\documentclass[prd,twocolumn,showpacs,amsmath,amssymb]{revtex4}
\usepackage{graphicx}
\usepackage{dcolumn}
\usepackage{bm}
\begin{document}
\title{Dynamical behaviors of FRW Universe containing a positive/negative potential scalar field in loop quantum cosmology}
\author{Xiao Liu}
 \email{lxiao@mail.bnu.edu.cn}
  \affiliation{Department of Physics, Beijing Normal University, Beijing 100875, China}
\author{Kui Xiao}
 \email{87xiaokui@mail.bnu.edu.cn}
  \affiliation{Department of Basic Teaching, Hunan Institute of Technology, Hengyang 421002x, China}
\author{Jian-Yang Zhu}
\thanks{Author to whom correspondence should be addressed}
 \email{zhujy@bnu.edu.cn}
  \affiliation{Department of Physics, Beijing Normal University, Beijing 100875, China}
\date{\today}

\begin{abstract}
The dynamical behaviors of FRW Universe containing a posivive/negative potential scalar field in loop quantum cosmology scenario are discussed. The method of the phase-plane analysis is used to investigate the stability of the Universe. It is found that the stability properties in this situation
are quite different from the classical cosmology case. For a positive potential scalar field coupled with a barotropic fluid, the cosmological autonomous system has five fixed points and one of them is stable if the adiabatic index $\gamma$ satisfies $0<\gamma<2$. This leads to the fact that the universe just have one bounce point instead of the singularity which lies in the quantum dominated area and it is caused by the quantum geometry effect. There are four fixed points if one considers a scalar field with a negative potential, but none of them is stable. Therefore, the universe has two kinds of bounce points, one is caused by the quantum geometry effect and the other is caused by the negative potential, the Universe may enter a classical re-collapse after the quantum bounce. This hints that the spatially flat FRW Universe containing a negative potential scalar field is cyclic.
\end{abstract}

\maketitle

\section{Introduction}
The cyclic model of universe was first noticed by Einstein who considered the possibility of this model as an alternative to the model of an expanding universe. But this model is failed for the entropy problem \cite{Tolman-1934}. Since then, many cyclic universe models had been introduced, i.e., the Brane cosmology model \cite{Steinhardt-SCI,Steinhardt-2002}, the Baum-Frampton model \cite{Baum-2007}, cyclic universe caused by loop quantum gravity (LQG) \cite{Bojowald-cyc,H.H.Xiong-cyc,kevin-cyc,kevin-cyc2,antide-cyc}, etc. The universe may enter a cyclic stage if it contains a scalar field with a negative potential \cite{Heard-negative,Steinhardt-2005}. Considering the flat Friedmann-Robertson-Walker(FRW) universe, the Friedmann equation reads
\begin{eqnarray}
  H^2=\frac{\kappa}{3}\left(\frac12\dot{\phi}^2+V(\phi)\right)\label{Fri-C},
\end{eqnarray}
with $\kappa=8\pi G$, and $\phi$ is the scalar field. Obviously, if the potential $V(\phi)$ is negative, the universe may enter a re-collapse phase, i.e., the Hubble parameter $H=0$, and then $\dot{a}=0$ if $\frac12\dot{\phi}^2=-V(\phi)$ with $V(\phi)<0$. But there are other possibilities that the universe will not enter the cyclic phase because the quantum singularity can not be avoided sometimes. To avoid the quantum singularity, we consider the universe described by loop quantum cosmology (LQC).

LQC \cite{Bojowald-Living,Ashtekar-Report} is a canonical quantization of homogeneous spacetime based on the techniques used in loop quantum gravity (LQG) \cite{Rovelli-Book,Thiemann-Book}. In the present isotropic and homogeneous settings, the canonically conjugate fields of LQG are parameterized by constants $c$ and $p$ respectively\cite{Ashtekar-Report}:
\begin{equation}
A^i_a=c\,\,l^{-1}_0\,{^0\omega}^i_a,\,\,and\,\,E^a_i=p\,\,l^{-2}_0 \sqrt{^0q}\,\,{^0e}^a_i.
\end{equation}
It has been shown that the loop quantum effects can be very well described by an effective modified Friedmann dynamics. Two corrections of the effective LQC are always considered: the inverse volume correction and the holonomy correction. But the inverse volume modification suffers from gauge dependence which cannot be cured and thus yields unphysical effects. Moreover, for a Universe with a large scale factor, the inverse volume modification to the Friedmann equation can be neglected and only the holonomy modification is important.

In this paper, we will use phase-plane analysis of the cosmological autonomous system to investigate the dynamical behavior of the FRW universe with a positive/negative potential scalar field (see the Eq. (\ref{potential})) in effective LQC based only on the holonomy modification. By this, the Friedmann equation adds a $-\frac{\kappa}{3}\frac{\rho^2}{\rho_c}$ term to the right-hand side of the standard Friedmann equation \cite{Ashtekar-improve}. Since this
correction comes with a negative sign, the Hubble parameter $H$, and then $\dot{a}$ will vanish when $\rho=\rho_c$, and the quantum bounce occurs.  Based on the holonomy modification, the dynamical behavior of dark energy has recently been investigated by many authors \cite{Wei-dy,Fu-Yu-Wu,Li-Ma,Lamon-Woeh,Samart-Phantom,Xiao-dynamical}. The attractor behavior of scalar field in LQC has also been studied \cite{Xiao-PRD,Copeland-superinflation,Lidsey-attractor}. The dynamical behavior of scalar field with positive/negative potential(Eq. (\ref{potential})) in classical cosmology has been discussed by \cite{Heard-negative}.

The organization of this paper is as follows. In the Sec. \ref{SEC2}, we present the general equations of motion (EOM). In the Sec. \ref{SEC3}, we introduce the autonomous system and analyse the dynamical behavior of this system. The numerical results will be presented in the Sec. \ref{SEC4}. Finally, the conclusions will be devoted in the Sec. \ref{SEC5}.

\section{Basic equations}\label{SEC2}
We focus on the flat FRW cosmology. The modified Friedmann equation in the effective LQC with holonomy correction can be written as \cite{Ashtekar-improve}
\begin{eqnarray}\label{Feq}
H^2=\frac{1}{3}\rho\left(1-\frac{\rho}{\rho_c}\right),\label{Fri}
\end{eqnarray}
in which $\rho$ is the total energy density and the natural unit $\kappa=8\pi G=1$ is adopted for simplicity. We consider a self-interacting scalar field $\phi$ with a potential $V(\phi)$ (for a more general consideration of self-interacting scalar field theory in the universe with holonomy effects resulting from the Loop Quantum Cosmology, see \cite{Mielczarek-PLB}) coupled with a barotropic fluid. Then the total energy density can be written as $\rho=\rho_\phi+\rho_\gamma$, with the energy density of scalar field $\rho_\phi=\frac12\dot{\phi}^2+V(\phi)$ and the energy density of barotropic fluid $\rho_\gamma$. It is worth pointing out that for the case of negative potential if the energy density of the scalar field remains positive, $\dot{\phi}$ must satisfying $\dot{\phi}^2>-V(\phi)$.

We consider a simple exponential potential for the scalar field $\phi$
\begin{equation}
V(\phi)=V_0e^{-\lambda\phi}\label{potential},
\end{equation}
where $\lambda$ is a dimensionless constant characterizing the slope of the potential \cite{Heard-negative}, and $V_0$ can be a positive or a negative constant corresponding to positive or negative potential. The positive/negateive potential has been discussed by many authors in classical cosmology \cite{Heard-negative} and in loop quantum cosmology. If $V_0>0$, the energy density of scalar field $\rho$ is always positive, and the Hubble constant $H$ will be zero if and only if $\rho=\rho_c$, which is just the quantum bounce caused by the quantum geometry effect. But if $V_0<0$, it is obvious that the Hubble parameter $H$ can be zero if $\rho=0$, i.e., $\rho_\gamma+\frac12\dot{\phi}^2=-V(\phi)$. This means that the universe has one quantum bounce phases when $\rho=\rho_c$, and one re-collapsing phase lying in the classical stage when $\rho_\gamma+\frac12\dot{\phi}^2=-V(\phi)$.

We consider that the energy momentum of this field to be covariantly conserved. Then one has
\begin{eqnarray}
&&\ddot{\phi}+3H\dot{\phi}+V'=0,\label{ddotphi}\\
&&\dot{\rho}_\gamma+3\gamma H\rho_\gamma=0, \label{dotrg}
\end{eqnarray}
where $\gamma$ is an adiabatic index that satisfies $p_\gamma=(\gamma-1)\rho_\gamma$ with $p_\gamma$ being the pressure of the barotropic fluid, and the prime denotes the differentiation w.r.t. the field $\phi$. Differentiating Eq. (\ref{Fri}), and using Eqs. (\ref{ddotphi}) and (\ref{dotrg}), one can obtain
\begin{eqnarray}
\dot{H}=-\frac12\left(\dot{\phi}^2+\gamma\rho_\gamma\right)\left[1-\frac{2(\rho_\gamma+\rho_\phi)}{\rho_c}\right].
\label{Fri4}
\end{eqnarray}
Considering the situation of $V_0<0$, obviously, $\dot{H}>0$ when $\rho=\rho_c$, while $\dot{H}<0$ when $\rho_\gamma+\frac12\dot{\phi}^2=-V(\phi)$. Let¡¯s assume the universe is expanding at the beginning without losing generality, then the scalar factor $a$ is increasing and the total energy density $\rho$ is decreasing. When $\rho=0$, i.e., $\frac12\dot{\phi}^2+\rho_\gamma=-V(\phi)$, $H=0$ and $\dot{H}<0$, the turnaround happens. The universe ceases expanding and turns to a contracting phase. In the contracting phase, the scale factor $a$ is decreasing, while $\rho$ is increasing. In this phase, if $\rho=\rho_c$, $H=0, \dot{H}>0$, the bounce occurs, and the universe reenters the expanding phase. To conclude, a scalar field with negative potential in loop quantum cosmology can evolve cyclically. We will show these pictures in the Sec.\ref{SEC4}. Noted that the analysis in this paper is based on effective LQC. This is different with \cite{antide-cyc}, which is based on the quantum theory of LQC.

\section{Dynamical Analysis}\label{SEC3}
In the last section, we introduced the basic equations. From Eq. (\ref{Feq}), we can see that a new term is added to the Friedmann equation in effective LQC. Based on the research in \cite{Xiao-PRD}, this lead to the consequence that the dynamical system will be very different from the one in classical cosmology \cite{Heard-negative,Copeland-exponential}, i.e., one additional dimension will appear in the autonomous system of LQC. So, the stability properties of the universe will change. It is interesting to see if there exists any stable point in the new autonomous system and checking out whether the result is consistent with our prediction presented in the Sec. \ref{SEC2}, because if the autonomous system contain a stable point, the universe will finally evolve into a steady state and will not cyclic. This discussion also can help us to to understand how LQC changes the conclusions obtained in classical situation. In this and the next section, we will discuss the dynamical behavior of the universe containing a scalar field with positive/negative potential.

Equations (\ref{Fri})-(\ref{ddotphi})  characterize a closed system which can determine the cosmic behavior. We introduce three dimensionless variables
\cite{Copeland-exponential}
\begin{equation}
x\equiv\frac{\dot{\phi}}{\sqrt{6}H},\,\,\,y\equiv\frac{\sqrt{|V|}}{\sqrt{3}H},\,\,\,z\equiv\frac{\rho}{\rho_c}.\label{xyz}
\end{equation}
Then the Friedmann equation (see Eq.(\ref{Fri})) can take the simple form
\begin{equation}
\left(x^2\pm y^2+\frac{\rho_\gamma}{3H^2}\right)(1-z)=1,\label{simh}
\end{equation}
and one can get
\begin{equation}
\frac{\rho_\gamma}{3H^2}=\frac{1}{1-z}-x^2\mp y^2.\label{simh2}
\end{equation}
We use upper/lower signs to denote the two distinct cases of $\pm V>0$ here. The $z$ term is a special variable in LQC.  $0<z<1$ when the Universe lies in quantum dominated stage, i.e., $\rho\lesssim\rho_c$, and $z=0$ when the universe lies in classical stage, i.e., $\rho\ll\rho_c$. Differentiating Eq. (\ref{simh}) one can obtain
\begin{equation}
\frac{\dot{H}}{H^2}=-\left[3x^2+\frac{3\gamma}{2}\left(\frac{1}{1-z}-x^2\mp y^2\right)\right](1-2z).\label{doth}
\end{equation}

Based on the above dimensionless variables (\ref{xyz}), we will use the phase-plane analysis of the cosmological autonomous system to investigate the dynamical behavior of this exponential potential scalar field coupled with a barotropic fluid in the next two subsections and show some numerical analysis in the next section.

\subsection{Autonomous system}

Using the new variables (\ref{xyz}), and considering Eq. (\ref{simh2}) and Eq. (\ref{doth}), we can rewrite Eqs. (\ref{Fri})-(\ref{ddotphi}) in the flowing form
\begin{widetext}
\begin{eqnarray}
\frac{dx}{d
N}&=&-3x+x\left[3x^2+\frac{3\gamma}{2}(\frac{1}{1-z}-x^2\mp y^2)\right]\times(1-2z)\pm\sqrt{\frac{3}{2}}\lambda y^2,\label{x'}\\
\frac{dy}{d N}&=&y\left[3x^2+\frac{3\gamma}{2}(\frac{1}{1-z}-x^2\mp y^2)\right]\times(1-2z)-\sqrt{\frac{3}{2}}\lambda x y,\label{y'}\\
\frac{dz}{d N}&=&2z(1-z)\left[-3x^2-\sqrt{\frac{3}{2}}\lambda\right. xy^2\pm\sqrt{\frac{3}{2}}\lambda x y^2\left.-\frac{3\gamma}{2}(\frac{1}{1-z}-x^2\mp y^2)\right],\label{z'}
\end{eqnarray}
where $N=\ln a$.
\end{widetext}

Obviously, the terms on the right-hand side of  Eqs. (\ref{x'})-(\ref{z'}) only depend on $x,y,z$ but not on $N$ or other variables. Such a dynamical system is usually called an autonomous system. We will restrict our discussion of the existence and stability of critical points of this autonomous system to the upper
half-plane $y\geq0$, consider only $\lambda\geq0$ without loss of generality.

This autonomous system is 3-dimensional for we choose the exponential potential (\ref{potential}). If one do not consider this special potential, it is possible to get one or more extra dimensions \cite{Xiao-PRD}.

\subsection{Stability properties}
Now we consider the fixed points of this autonomous system described by Eqs. (\ref{x'})-(\ref{z'}). Let $f=dx/dN,g=dy/dN,h=dz/dN$. The fixed points of the autonomous system satisfy the condition
\begin{equation}
(f,g,h)|_{x_c,y_c,z_c}=0.
\end{equation}
According to this condition, it is easy to get the values of the fixed points, and they are shown in the Table I.

The properties of each fixed point are determined by the eigenvalues of the Jacobi matrix
\begin{eqnarray}
{\cal{M}}= \left. \begin{pmatrix}
\frac{\partial f}{\partial x}&\frac{\partial f}{\partial y}&\frac{\partial f}{\partial z}\\
\frac{\partial g}{\partial x}&\frac{\partial g}{\partial y}&\frac{\partial g}{\partial z}\\
\frac{\partial h}{\partial x}&\frac{\partial h}{\partial y}&\frac{\partial h}{\partial z}\\
\end{pmatrix}
\right|_{(x_c,y_c,z_c)}.
\end{eqnarray}

According to Lyapunov¡¯s linearization method, the stability of a linearized system is determined by the eigenvalues of the matrix $\cal{M}$. If all of the eigenvalues are strictly in the left-half complex plane, then the autonomous system is stable. If at least one eigenvalue is strictly in the right-half complex plane, then the system is unstable.

\begin{widetext}
\begin{center}
\begin{table}[!ht]
\caption{The stability analysis of an autonomous system in LQC. The system is described by a self-interacting scalar field $\phi$ with exponential potential $V$ coupled with a barotropic fluid $\rho_\gamma$. Explanation of the symbols used in this table: $P_{i}$ denotes the fixed points located in the 3-dimensional phase space, which are marked by the coordinates $(x_c,y_c,z_c,\lambda_c)$. $\bf{M}^T$ means the inverted matrix of the eigenvalues of the
fixed points. U stands for unstable, and S stands for stable.}
\begin{ruledtabular}
\begin{tabular}[c]{cccccccc}
Fixed-points & $x_c$ & $y_c$&$z_c$ & Existence  & Eigenvalues & Stability \\
\hline
$P_1$&$0$&$0$&$0$& All $V$,$\lambda$ and $\gamma$ & $\bf{M}^T=(-3+\frac32\gamma,-3\gamma,\frac32\gamma)$& U, for all $\lambda,\gamma$\\
\hline
$P_2$&$1$&$0$&$0$ &All $V$,$\lambda$ and $\gamma$ &  $\bf{M}^T=(-6,6-3\gamma,3-\frac{\sqrt{6}}{2}\lambda)$& U, for all $\lambda,\gamma$\\
\hline
$P_3$&$-1$&$1$&0&All $V$,$\lambda$ and $\gamma$ &$\bf{M}^T=(-6,6-3\gamma,3+\frac{\sqrt{6}}{2}\lambda)$& U, for all $\lambda,\gamma$\\
\hline
$P_{4_+}$&$\frac{\lambda}{\sqrt{6}}$&$\sqrt{1-\frac{\lambda^2}{6}}$&$0$& $V>0$ and $\lambda^2<6$ &$\bf{M}^T=(-\lambda^2,-3+\frac{\lambda^2}{2},\lambda^2-3\gamma)$&S, for $\lambda^2<3\gamma$\\
\hline
$P_{4_-}$&$\frac{\lambda}{\sqrt{6}}$&$\sqrt{\frac{\lambda^2}{6}-1}$&$0$& $V<0$ and $\lambda^2<6$ &$\bf{M}^T=\left(\lambda^2-\frac{\lambda^4}{3},-3+\frac{\lambda^2}{2},\lambda^2-3\gamma\right)$&U, for all $\lambda,\gamma$\\
\hline
$P_5$&$\frac{\sqrt{6}\gamma}{2\lambda}$&$\sqrt{\frac{3\gamma}{2\lambda^2}(2-\gamma)}$&$0$&$V>0$ and $\lambda^2>3\gamma$&See Eq.(\ref{P5}) &U, for all $\lambda,\gamma$
\end{tabular}
\end{ruledtabular}
\end{table}
\end{center}
\end{widetext}

Remarks on Tab.\textbf{I}:

(i). In fact, for $V<0$, we have another fixed point
\begin{eqnarray}
(x_c,y_c,z_c)=\left(\frac{\sqrt{6}}{2\lambda}\gamma,\sqrt{\frac{3\gamma}{2\lambda^2}(\gamma-2)},0\right),\nonumber
\end{eqnarray}
however, for simplicity, we just consider $0<\gamma<2$. This means that $y_c$ is a complex number and is thus a non-physical solution. Since the adiabatic index $\gamma$ satisfies $0<\gamma< 2$, all the terms that contain $\gamma$ should not change sign. If one consider $\gamma=0$ or $\gamma=2$, some of the eigenvalues corresponding to points P$_{1,2,3,5}$ will be zero. To analyze the stability of such points with zero eigenvalues, we need to resort to other more complex methods \cite{Xiao-PRD}.

(ii). The eigenvalues of the Point P$_5$ is
\begin{eqnarray}\label{P5}
{\cal{M}}=\begin{pmatrix}-3\gamma\\  \frac34{\frac {-2 \lambda+\gamma\lambda+\sqrt {4{\lambda}^{2}-20{\lambda}^{2}\gamma+9{\lambda}^{2}{\gamma}^{2}-24{\gamma}^{3}+48{\gamma}^{2}}}{\lambda}}\\
\frac34{\frac{-2\lambda+\gamma\lambda-\sqrt{4{\lambda}^{2}-20{\lambda}^{2}\gamma+9{\lambda}^{2}{\gamma}^{2}-24{\gamma}^{3}+48{\gamma}^{2}}}{\lambda}}
\end{pmatrix}.
\end{eqnarray}
Obviously, $4{\lambda}^{2}-20{\lambda}^{2}\gamma+9{\lambda}^{2}{\gamma}^{2}-24{\gamma}^{3}+48{\gamma}^{2}\geq 0$ should be held. This point is unstable because $F_1=-2 \lambda+\gamma\lambda+\sqrt{4{\lambda}^{2}-20{\lambda}^{2}\gamma+9{\lambda}^{2}{\gamma}^{2}-24{\gamma}^{3}+48{\gamma}^{2}}<0$ and  $F_2=-2 \lambda+\gamma\lambda-\sqrt{4{\lambda}^{2}-20{\lambda}^{2}\gamma+9{\lambda}^{2}{\gamma}^{2}-24{\gamma}^{3}+48{\gamma}^{2}}<0$ can not be satisfied at the same time. It is also impossible that $F_1=F_2=0$, since this means that $\gamma=2$. We do not consider this condition.

Based on Table I and the remarks above, we can find that there is no stable point when one consider the scalar field with a negative exponential potential. There is just one stable point (the Point P$_{4_+}$) when $\lambda^2<3\gamma$ and $\lambda^2<6$, if one consider the scalar field with a positive exponential potential. This conclusion is different from the stability properties of the scalar field with positive exponential potential in classical cosmology \cite{Copeland-exponential}.

We find the values of coordinates $(x_c,y_c)$ of P$_{4_+}$ and the fourth fixed point in the Table I of \cite{Copeland-exponential}, which is in the classical setting, are the same. So are their stability properties. We also can find there exists one more stale point, i.e., the fifth fixed point in the Table I of \cite{Copeland-exponential}. It has the same values of fixed point with P$_{5}$, but obviously, in our case, P$_5$ is unstable. This difference comes from the modification of Friedmann equation. Comparing the classical Friedmann equation Eq.(\ref{Fri-C}) and the modified Friedmann equation Eq.(\ref{Fri}), we can find that the modified Friedmann equation adds a $-\frac13\frac{\rho^2}{\rho_c}$ term, and then the dimension of the autonomous system in LQC is different from the one in classical cosmology.

\section{Numerical Results}\label{SEC4}
In this section, we will analyze the dynamical behavior of FRW universe containing a scalar field with positive/negative potential by using numerical tool.

\begin{figure}[!ht]
\includegraphics[clip,width=0.45\textwidth]{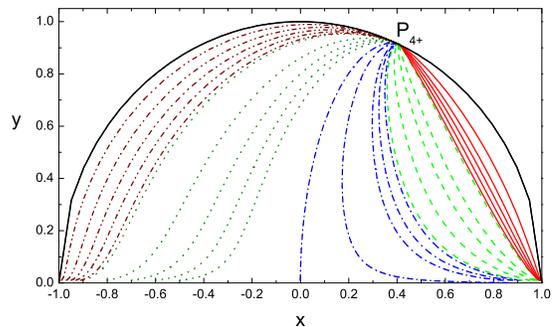}
\caption{The phase picture of the positive potential projection on the $x-y$ plane. The barotropic fluid is radiation, i.e., $\gamma=\frac43$. }
\label{fig1}
\end{figure}

We found that there is not any stable fixed points if one considers a scalar field with a negative exponential potential, and there is only one stable fixed point if one discuss a scalar field with a positive exponential potential. We presented this stable fixed point in Fig.\ref{fig1}. As it shows, almost all solutions of the dynamical equations of the autonomous system end up at $P_{4_+}$.

The evolution pictures of $H(t),a(t),\phi(t),\rho(t),\dot{\phi}(t)$ and $\rho_\gamma(t)$ are shown in the Fig.\ref{fig2}. We consider the time $t=0$ is the quantum bounce point. Then one has $H(t=0)=0$. The initial conditions of $\phi,\rho_\gamma$ and $a$ are $\phi(t=0)=\frac12, \rho_\gamma(t=0)=0.01$ and $a(t=0)=0.1$ for both the universe containing a scalar field with positive potential and the one with negative potential. According to these initial conditions and the Friedmann equation (see the Eq. (\ref{Fri})), one can get the initial condition of $\dot{\phi}$. It is easy to find that the spatially flat FRW universe is cyclic if we consider the scalar field with a negative exponential potential, but not cyclic if one consider the positive exponential potential. We have to point out that the initial conditions do not affect the cyclic behavior of the universe, different initial condition only change the locations of the fixed points and the universe evolution rate.

Here, we analyze these evolution pictures focusing on the situation of negative exponential potential. Fig.2(a)-(c) show that the evolution of the Hubble parameter $H(t)$, the scale factor $a(t)$ and the total energy density $\rho$. Just like we pointed out before, there are two sufficient conditions to insure $H(t)$ can be zero. The one takes place in the contracting phase, where the scale factor is decreasing, total energy density is increasing, and when $\rho=\rho_c$, $H=0$, the bounce occurs. Then the universe enters into an expanding phase. Different from the classical cyclic universe, with loop quantum corrections, the universe bounces back before a singularity occurs. The other one occurs when $\rho=0$, which means $\frac{1}{2}\dot{\phi}^2+V(\phi)+\rho_{\gamma}=0$. Because the total energy density equals to zero, this condition lies in the classical stage. From the image, we see this corresponds to the zero points where $H(t)$ evolves from positive to negative, $a(t)$ evolves from increasing to decreasing, and $\rho(t)$ evolves from decreasing to increasing.  From Fig.2(d) and Fig.2(e), we can see that the scalar field is not cyclic, and its momentum continuously decreases. This consequence comes from the modified Friedmann equation. Combined with Fig.2(c) and Fig.2(f), we can find that $\dot{\phi}(t)$ can not reach exactly zero, so that $\phi(t)$ will not turn around but will continue to increase and just slow down during the bounce and re-collapse. The bounce and re-collapse in scale factor and Hubble parameter are not the bounce and re-collapse in $\phi(t)$. So there is a series of bounces and re-collapses at finite values of $\phi(t)$. Meanwhile, since the potential is not constant, the behavior of bounce and re-collapse of the universe is not periodic but still cyclic \cite{Bojowald-cyc}. Just as Fig.2(a) shows, the cycle period of $H(t)$ is rising gradually.

\begin{widetext}
\begin{center}
\begin{figure}[!ht]
\includegraphics[clip,width=0.85\textwidth]{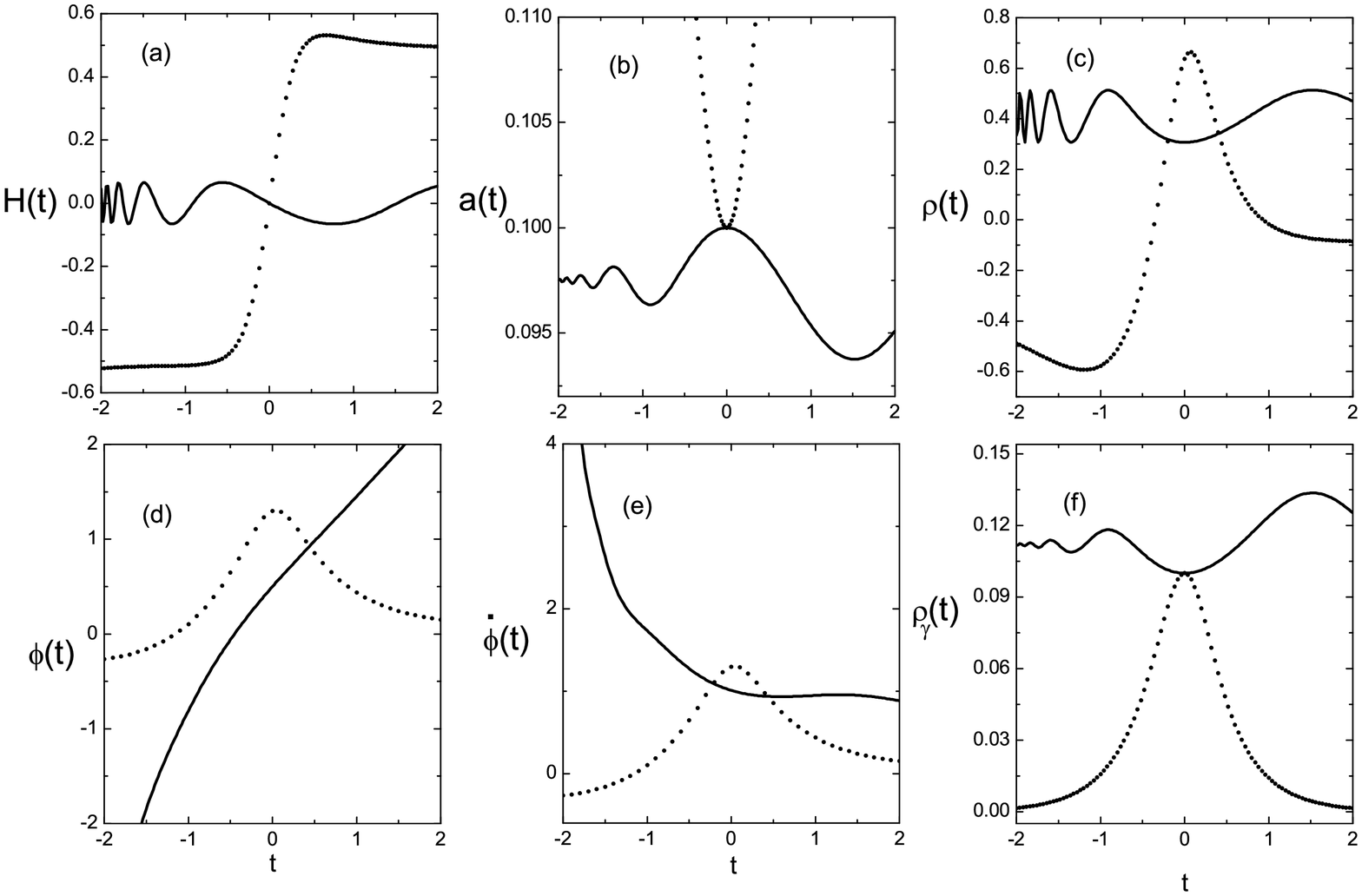}
\caption{The evolution pictures of the universe containing a positive potential/negative potential scalar field, for representing (a) $H(t)$, (b) $a(t)$, (c) $\rho(t)$, (d) $\phi(t)$, (e) $\dot{\phi}(t)$ and (f) $\rho_\gamma(t)$, respectively. The initial conditions are $H(t=0)=0$, $a(t=0)=0.1$, $\rho_\gamma(t=0)=0.01$, $\phi(t=0)=\frac12$, and $\dot{\phi}(t=0)=1.4922$ for positive potential (the pointed line), and $H(t=0)=0$, $\rho_\gamma(t=0)=0.01$, $\phi(t=0)=\frac12$, and $\dot{\phi}(t=0)=1.007$ for negative potential (the solid line). The barotropic fluid is radiation, i.e., $\gamma=\frac43$.}.  \label{fig2}
\end{figure}
\end{center}
\end{widetext}

\section{Conclusions}\label{SEC5}
The goal of this paper is to discuss the dynamical properties of the FRW universe with a positive/negative exponential potential scalar field coupled with a barotropic fluid in LQC scenario.

The Friedmann equation is modified in LQC scenario. In this situation the universe will enter a quantum bounce phase when the total energy density $\rho$ is equal to the critical energy density $\rho_c$. After quantum bounce, the universe will immediately enter a super-inflation phase, and then enter a stage which is described very well by general relativity. According to the phase-plane analysis of the evolution of the spatially flat FRW universe in LQC scenario, we found that there is just one stable fixed point (see the Point P$_{4_+}$ in the Table I). So, if the universe is described by flat FRW metric, just including the scalar field with positive potential and coupled with a barotropic fluid, it has only one quantum bounce. This is different from the stability properties of classical cosmology which includes the same fields \cite{Copeland-exponential}. The difference comes from the modification of Friedmann equation. Since the Friedmann equation is modified in LQC scenario, the autonomous system is 3-dimensional in LQC, while 2-dimensional in classical cosmology.

Also, we have presented a phase-plane analysis of the evolution of a spatially flat FRW universe containing a scalar field with a negative exponential potential $V=V_0 \exp(-\lambda\phi)$ $(V_0<0)$ plus a barotropic fluid. Because of the different potential, the dynamical equation and thus the dynamical behavior of these two situations are different. When we consider the situation of the scalar field with positive exponential potential, the autonomous system has five fixed points, and one of them is stable if one constrains the adiabatic index with $0<\gamma<2$. But when the potential is negative, there exist just four fixed points, and none of them is stable under the same $\gamma$. The universe may enter a classical re-collapse after the quantum bounce in this scenario if $\frac12\dot{\phi}^2+\rho_\gamma=-V_0\exp(-\lambda\phi)$. The spatially flat FRW Universe is cyclic.

To summarize, from the mathematical point of view, the necessary condition for the universe become cyclic is that there is no stable point. In the model we discussed, just because the positive potential case does have one stable point, almost all solutions of the dynamical equations of the autonomous system end up at the stable point. This means finally, the universe will evolve into a very steady state, rather than a cyclic stage. Meanwhile the negative exponential potential is not the only potential that can lead to cyclic universe. For a cyclic universe model caused by another kind of scalar field potential in LQC, see \cite{kevin-cyc}, while \cite{kevin-cyc2} considered the situation where the space is anisotropic.

\acknowledgments This work was supported by the National Natural Science Foundation of China (Grant Nos. 11175019 and 11235003) and Xiao was also supported by the National Natural Science Foundation of China (Grant No. 11247282).

\end{document}